\documentclass[11pt]{article}
\usepackage[a4paper]{geometry}
\usepackage{array}
\usepackage{amssymb}
\usepackage{amsfonts}
\usepackage{amsmath}
\usepackage{mathrsfs}
\usepackage{latexsym}
\usepackage{graphicx}
\usepackage{subfigure}
\usepackage{psfrag}
\usepackage{enumerate}
\usepackage{url}
\usepackage{color}
\usepackage{nicefrac}
\usepackage{verbatim}
\usepackage{authblk}
\usepackage{textcomp}
\usepackage{tikz}
\usetikzlibrary{fit,calc,positioning,decorations.pathreplacing}

\def\mkvc{\textsc{max $k$-vertex cover}}

\def\opt{\mathrm{opt}}
\def\SOL{\mathrm{SOL}}

\newtheorem{THEOREM}{Theorem}
\newenvironment{theorem}{\begin{THEOREM} \hspace{-.85em} {\bf .} }%
                        {\end{THEOREM}}
\newtheorem{LEMMA}{Lemma}
\newenvironment{lemma}{\begin{LEMMA} \hspace{-.85em} {\bf .} }%
                      {\end{LEMMA}}
\newtheorem{CLAIMEMPH}{Claim}
\newenvironment{claimemph}{\begin{CLAIMEMPH} \hspace{-.85em} {\bf .} }%
                            {\end{CLAIMEMPH}}

\newcommand{\thm}{\begin{theorem}}
\newcommand{\lem}{\begin{lemma}}
\newcommand{\clme}{\begin{claimemph}}
\newcommand{\prf}{\noindent{\bf Proof.} }
\newcommand{\ethm}{\end{theorem}}
\newcommand{\elem}{\end{lemma}}
\newcommand{\eclme}{\end{claimemph}}
\newcommand{\eprf}{\bbox}
\newcommand{\bbox}{\vrule height7pt width4pt depth1pt}

\title{\textbf{Combinatorial approximation of maximum $k$-vertex cover in bipartite graphs within ratio~0.7}}
\author{Vangelis~Th.~Paschos\footnote{Also, Institut Universitaire de France} \\ 
PSL* Research University, Universit\'e Paris-Dauphine, LAMSADE \\
CNRS UMR 7243 \\
\texttt{paschos@lamsade.dauphine.fr}
}

\begin{document}

\maketitle

\begin{abstract}

We propose a \textit{purely combinatorial algorithm} for \mkvc{} in bipartite graphs, achieving approximation ratio~0.7. The only combinatorial algorithms currently known until now for this problem  are the natural greedy algorithm, that achieves ratio $\nicefrac{(e-1)}{e} = 0.632$, and an easy~$\nicefrac{2}{3}$-approximation algorithm presented in~\cite{DBLP:journals/corr/BonnetEPS14}.

\end{abstract}

\section{Introduction}\label{intro}

In the \mkvc{} problem, a graph~$G=(V,E)$ with $|V| = n$ and $|E| = m$ is given together with an integer $k \leqslant n$. 
The goal is to find a subset $K \subseteq V$ with $k$ elements such that the total number of edges covered by~$K$ is maximized. 
This problem is \textbf{NP}-hard even in bipartite graphs~\cite{apollonio14,DBLP:conf/ifipTCS/CaskurluMPS14}.

The approximation of \mkvc{} has been originally studied in~\cite{Hochbaum98}, where an approximation $1-\nicefrac{1}{e}$ was proved, achieved by the natural greedy algorithm. This ratio is tight even in bipartite graphs~\cite{DBLP:conf/compgeom/BadanidiyuruKL12}.  In~\cite{ageev}, using a sophisticated linear programming method, the approximation ratio for \mkvc{} is improved up to~$\nicefrac{3}{4}$. Finally, by an easy reduction from \textsc{Min Vertex Cover}, it can be shown that \mkvc{} in general graphs does not admit a polynomial time approximation schema (PTAS), unless $\mathbf{P} = \mathbf{NP}$~\cite{DBLP:journals/cc/Patrank94}.

Obviously, the result of~\cite{ageev} immediately applies to the case of bipartite graphs. Very recently,~\cite{DBLP:conf/ifipTCS/CaskurluMPS14} has improved this ratio in bipartite graphs up to~$\nicefrac{8}{9}$, always using involved linear programming techniques, but the existence of a~PTAS for such graphs always remains open.

Finally, let us note that \mkvc{} is polynomial in regular bipartite graphs or in semi-regular ones, where the vertices of each color class have the same degree. Indeed, in both cases it suffices to chose~$k$ vertices in the color class of maximum degree.

Our principal motivation for this paper is to study \textit{in what extent combinatorial methods for \mkvc{} compete with linear programming ones}. In other words, \textit{what is the ratios' level, a purely combinatorial algorithm can guarantee?} In this purpose, we devise an algorithm that builds five distinct solutions and returns the best among them; for this algorithm, we prove a worst case 0.7-approximation ratio. Let us note that a similar issue is presented in~\cite{Trevisan:2009:MCS:1536414.1536452} for \textsc{max cut} where a 0.531-ratio combinatorial algorithm is given. Comparison of classes of methods with respect to their abilities to solve problems seems to be a very interesting research issue. This may bring new insights to both the problems handled and the methods themselves. Furthermore, such studies may exhibit interesting and funny mathematical problems.

Note finally that in~\cite{DBLP:journals/corr/BonnetEPS14}, an easy~$\nicefrac{2}{3}$-approximation algorithm is presented, together with a very complex one where a computer assisted analysis was giving a ratio of~0.792. But this ratio is impossible to be proved analytically.

\section{Preliminaries}\label{prelim}

Consider a bipartite graph~$B(V_1,V_2,E)$, fix an optimal solution~$O$ for \mkvc{} (i.e., a vertex-set on~$k$ vertices covering a maximum number of edges in~$E$)  and denote by~$k_1$ and~$k_2$ the cardinalities of the subsets~$O_1$ and~$O_2$ of~$O$ lying in the color-classes~$V_1$ and~$V_2$, respectively. W.l.o.g., we assume $k_1 \leqslant k_2$ and we set:
\begin{eqnarray}
k_1 &=& \mu \cdot k_2, \;\;\; \mu \leqslant 1 \label{mu} \\
k &=& k_1 + k_2 \;\; = \;\; (1+\mu)\cdot k_2 \label{1+mu}
\end{eqnarray}
Denote by~$\delta(V')$, $V' \subseteq V = V_1 \cup V_2$, the number of edges covered by~$V'$ and by~$\opt(B)$ the value of an optimal solution (i.e., the number of edges covered by~$O$). 

\begin{figure}[h*]
\begin{center}
\includegraphics[width=\textwidth]{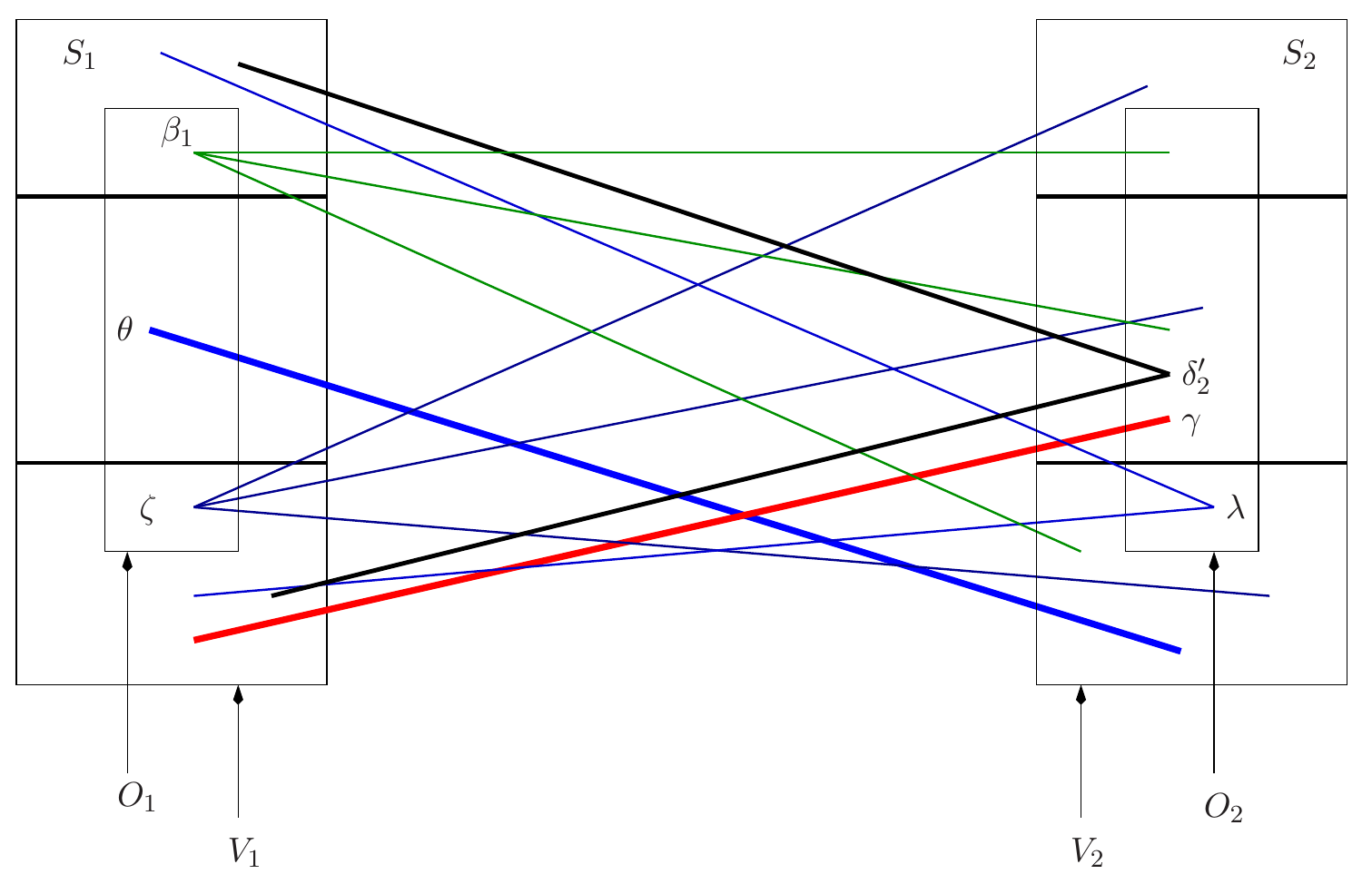}
\caption{The edge-sets induced by the several parameters.}\label{cuts}
\end{center}
\end{figure}

Let~$S_i$, $i = 1,2$, be the~$k_i$ vertices of~$V_i$ that cover the most of edges. Obviously,~$S_i$ is the set of the~$k_i$ largest degree vertices in~$V_i$ (breaking ties arbitrarily) and the following hold:
\begin{eqnarray}
\delta\left(S_1\right) &\geqslant& \delta\left(O_1\right) \label{delta1} \\
\delta\left(S_2\right) &\geqslant& \delta\left(O_2\right) \nonumber 
\end{eqnarray}
In what follows, we call ``best'' vertices, a set of vertices that cover the most of \emph{uncovered} edges\footnote{For instance, saying ``we take~$S_1$ plus the~$k_2$ best vertices in~$V_2$, this means that we take~$S_1$ and then~$k_2$ vertices of highest degree in~$B[(V_1 \setminus S_1),V_2]$.} in~$B$. Furthermore, we will also use the following additional notations and conventions (we assume that vertices in both~$V_1$ and~$V_2$ are ordered in decreasing degree order), where all the greek letters used imply parameters that are all smaller than, or equal to,~1:
\begin{itemize} 
\item $\delta(O_1)$: the number of edges covered by~$O_1$; for conciseness we set $\delta(O_1) = \alpha\cdot\opt(B)$; 
\item $\beta_1\cdot\delta\left(O_1\right) = \beta_1\cdot\alpha\cdot\opt(B)$: the number of edges covered by~$S_1 \cap O_1$; 
\item $\delta'(O_2)$: the number of \textit{private} edges covered by~$O_2$, i.e., the edges already covered by~$O_1$ are not counted up to~$\delta'(O_2)$; obviously, $\delta'(O_2) = (1-\alpha)\cdot\opt(B)$;
\item $\theta\cdot\delta(O_1)$: the number of edges (if any) from~$O_1$ that go ``below''~$O_2$ (recall~$V_1$ and~$V_2$ are ordered in decreasing degree order);
\item $\gamma\cdot\delta'(O_2)$: symmetrically, it denotes the number of edges of~$O_2$ that go below the vertices of~$O_1$;
\item $\zeta\cdot\delta(O_1)$: suppose that after taking the~$k$ best vertices of~$V_1$, there still remain, say,~$k'_1$ vertices of~$O_1$ that have not been encountered yet;~then,~$\zeta\cdot\delta(O_1)$ is the number of edges covered by those vertices;
\item $\lambda\cdot\delta'(O_2)$: this is the symmetric of the quantity~$\zeta\cdot\delta(O_1)$ for the pair~$(V_2,O_2)$ (supposing that the number of vertices in~$O_2$ that have not been encountered is~$k'_2$).
\end{itemize}
In Figure~\ref{cuts}, the edge-sets defined by the parameters above are illustrated. Heavy lines within rectangles~$V_1$ and~$V_2$ represent the borders of~$S_1$ and~$S_2$ (the upper ones) and those of the~$k$ best vertices (the lower ones). Edges from~$O_1$~($\arg(\delta(O_1))$) are not shown in the the figure. They can go everywhere in~$V_2$. Private edges of~$O_2$~($\arg(\delta'(O_2))$) are shown as heavy lack lines (the set of edges~$\delta'_2$). They can go everywhere in~$V_1 \setminus O_1$.

The basic idea of the algorithm is quite simple. It computes the best among five solutions built. It is presented in Section~\ref{approxalgsection} while,  in Section~\ref{approxratiosec}, we show that its approximation ratio is bounded below by~0.7. 

\section{The algorithm~\ldots}\label{approxalgsection}

The  algorithm guesses the cardinalities~$k_1$, $k_2$ of~$O_1$ and~$O_2$, respectively, builds the five solutions specified just below and returns the best among them:
\begin{itemize}
\item[] $\SOL_1$: take~$S_1$ plus the~$k_2$ remaining best vertices from~$V_2$;
\item[] $\SOL_2$: take~$S_2$ plus the~$k_1$ remaining best vertices from~$V_1$;
\item[] $\SOL_3$: take the~$k$ best vertices of~$V_2$;
\item[] $\SOL_4$: take the best between the following two solutions:
\begin{enumerate}
\item\label{kbest} the~$k$ best vertices of~$V_1$;
\item\label{2k1} the best~$2\cdot k_1$ vertices of~$V_1$ plus the remaining $k - 2\cdot k_1$ best vertices of~$V_2$.
\end{enumerate}
\end{itemize}
Let us note that the algorithm above, since it runs for any value of~$k_1$ and $k_2$, it will run for $k_1 = k$ and $k_2 = k$. So, it is optimal for the instances of~\cite{DBLP:conf/compgeom/BadanidiyuruKL12}, where the greedy algorithm attains the ratio~$\nicefrac{(e-1)}{e}$.

In what follows in this section, in Lemmata~\ref{sol1lem} to~\ref{sol4lem}, we analyze the solutions built by the algorithm and provide several expressions for the ratios achieved by each of them. All these ratios are expressed as functions of the parameters specified in Section~\ref{prelim}. In order to simplify notations from now on we shall write~$\opt$ instead of~$\opt(B)$.
\lem\label{sol1lem}
The approximation ratio achieved by solution~$\SOL_1$ is the maximum of the following quantities:
\begin{eqnarray}
1 - \alpha + \beta_1\cdot\alpha \label{ena} \\
\alpha + \gamma\cdot(1-\alpha) \label{tessera} 
\end{eqnarray}
Furthermore, if~$S_1$ and~$O_1$ coincide (i.e., $S_1 \cap O_1 = S_1$),~$\SOL_1$ is optimal.
\elem
\prf
For~(\ref{ena}),~$S_1$ covers, by~(\ref{delta1}), more than $\delta(O_1) = \alpha\cdot\opt$ edges. Decompose this edge-set into a set~$X$ of edges covered by $S_1 \setminus (S_1 \cap O_1)$ and the set of edges of size~$\beta_1\cdot\alpha\cdot\opt$ of edges covered by $S_1 \cap O_1$. On the other hand, the~$k_2$ remaining best vertices in~$V_2$ will cover more edges than the~$k_2$ remaining best vertices in~$O_2$, that cover more than $(1-\alpha)\cdot\opt - |X|$ edges, qed.

For~(\ref{tessera}), whenever~$S_1$ does not coincide with~$O_1$, there are vertices of~$O_1$ that ly below~$S_1$.  Since~$\gamma\cdot\delta'(O_2)$ is the number of edges from~$O_2$ that go belong~$O_1$, these edges will be not counted up in the set of edges covered by~$S_1$.

Finally, if~$S_1$ and~$O_1$ coincide,~$\SOL_1$ will cover $\alpha\cdot\opt + (1-\alpha)\cdot\opt = \opt$ edges.~\eprf
\lem\label{sol2lem}
The approximation ratio achieved by solution~$\SOL_2$ is bounded below by:
\begin{equation}
1 - \alpha + \alpha\cdot\theta \label{pente} 
\end{equation}
\elem
\prf
The proofs is similar with the one of Lemma~\ref{sol1lem} for~(\ref{tessera}).~\eprf
\lem\label{sol3lem}
The approximation ratio achieved by solution~$\SOL_3$ is the maximum of the following quantities:
\begin{eqnarray}
1 - \lambda\cdot(1-\alpha) - \alpha\cdot\theta \label{exi} \\
(1-\alpha)\cdot(1+\lambda\cdot\mu) \label{epta}
\end{eqnarray}
\elem
\prf
If after taking the~$k$ best vertices of~$V_2$ the whole of~$O_2$ has been encountered, all but~$\theta\cdot\delta(O_1)$ edges of the optimum have been covered. In this case, an appoximation ratio $1 - \alpha\cdot\theta$ is achieved. 

Otherwise, by the definition of~$\lambda\cdot\delta'(O_2)$:
$$
\opt - \lambda\cdot\delta'\left(O_2\right) - \theta\cdot\delta\left(O_1\right)  = \opt \cdot\left(1 - \lambda\cdot(1-\alpha) - \theta\cdot\alpha\right)
$$ 
edges of the optimum are covered.

On the other hand, taking the~$k$ best vertices of~$V_2$, consists of first taking~$S_2$ (covering $(1-\alpha)\cdot\opt$ edges) and then the~$k_1$ best vertices below it. Furthermore, below the~$k$ best vertices, the group of the~$k'_2$ ``worst'' vertices of~$O_2$ has average degree at least~$\nicefrac{\lambda\cdot\delta'(O_2)}{k'_2}$. Since the algorithm takes~$k_1$ ``better'' vertices, they will cover at least: 
$$
\frac{k_1}{k'_2}\cdot\lambda\cdot\delta'\left(O_2\right) \geqslant  \frac{k_1}{k_2}\cdot\lambda\cdot(1-\alpha)\cdot\opt {(\ref{mu}) \atop =} \mu\cdot\lambda\cdot(1-\alpha)\cdot\opt
$$
which proves~(\ref{epta}).~\eprf
\lem\label{sol4lem}
The approximation ratio achieved by solution~$\SOL_4$ is the maximum of the following quantities:
\begin{eqnarray}
1 - \zeta\cdot\alpha - \gamma\cdot(1-\alpha) \label{ennia} \\
\left(2 - \beta_1\right)\cdot\alpha + \frac{(1-\mu)\cdot\zeta\cdot\alpha}{\mu} \label{deka} \\
\frac{(1-\mu)+\alpha\cdot(2+\mu)-\alpha\cdot\beta_1}{2} + \frac{(1-2\cdot\mu)\cdot\zeta\cdot\alpha}{2\cdot\mu} \label{star}
\end{eqnarray}
\elem
\prf
Let us first note that, if after taking the~$k$ best vertices in~$V_1$ all the vertices of~$O_1$ are captured, the approximation ratio achieved is $1 - \gamma\cdot(1-\alpha)$ since only~$\gamma\cdot\delta'(O_2) = \gamma\cdot(1-\alpha)\cdot\opt$ edges of the optimum are not covered. Suppose now that~$k'_1$ verices of~$O_1$ are not captured. In this case, the~$k$ vertices taken from~$V_1$ cover: 
$$
\opt -\zeta\cdot\delta\left(O_1\right) - \gamma\cdot\delta'\left(O_2\right) = (1 - \zeta\cdot\alpha - \gamma\cdot(1-\alpha))\cdot\opt
$$
For~(\ref{deka}) and~(\ref{star}) now, observe first that the~$k$ vertices taken from~$V_1$ can be seen as the union of~$\nicefrac{k}{k_1}$ consecutive $k_1$-groups (called clusters in what follows) and that, by~(\ref{mu}) and~(\ref{1+mu}), $\nicefrac{k}{k_1} = \nicefrac{1+\mu}{\mu}$.
Assume also that the $k-k'_1$ of~$O_1$ encountered among the~$k$ best vertices of~$V_1$ are included in the~$\pi$ first clusters. Denote by~$\kappa_i$ the number of vertices of~$O_1$ in the~$i$-th cluster, $i= 1, \ldots, \pi$, and suppose that the ``optimal''~$\kappa_i$ vertices of cluster~$i$ cover $\beta_i\cdot \delta(O_1) = \beta_i\cdot\alpha\cdot\opt$ edges. 
\clme\label{fact1} 
\textsl{Consider cluster~$i$ and denote by~$\bar{O}_{1,i}$ the part of~$O_1$ not captured by clusters $1, 2, \ldots, i-1$ (so, $\bar{O}_{1,i} = \sum_{j = i}^{\pi}\kappa_j + k'_1$). Then, the vertices of cluster~$i$ will cover at least~$(1 - \sum_{j=1}^{i-1}\beta_j)\cdot\alpha\cdot\opt$ edges.}~\eclme
In order to prove Claim~\ref{fact1}, observe that the part of~$\delta(O_1)$ covered by~$\bar{O}_{1,i}$ is:
$$
\delta\left(\bar{O}_{1,i}\right) = \delta\left(O_1\right) - \sum_{j=1}^{i-1}\beta_j\cdot\delta\left(O_1\right) = \left(1 -\sum_{j=1}^{i-1}\beta_j\right)\cdot\alpha\cdot\opt
$$ 
and that the~$\delta(\bar{O}_{1,i})$ edges are covered by~$\sum_{j = i}^{\pi}\kappa_j + k'_1 = k_1 - (\sum_{j = 1}^{i-1}\kappa_j) \leqslant k_1$ vertices, while cluster~$i$ contains exactly~$k_1$ vertices with degree at least as large as those of~$\bar{O}_{1,i}$. An easy average argument derives then that the vertices of cluster~$i$ will cover at least:
$$
k_1\cdot\frac{\left(1 -\sum_{j=1}^{i-1}\beta_j\right)\cdot\alpha\cdot\opt}{k_1 - \left(\sum_{j = 1}^{i-1}\kappa_j\right)} \geqslant \left(1 -\sum_{j=1}^{i-1}\beta_j\right)\cdot\alpha\cdot\opt
$$
edges, qed.

Consider the two first groups clusters taken from~$V_1$. The first of them~($S_1$) covers more than $\delta(O_1) = \alpha\cdot\opt$ edges (by~(\ref{delta1}))  while, by Claim~\ref{fact1}, the second one will cover more than $(\nicefrac{k_1}{k_1 - \kappa_1})\cdot(1-\beta_1)\cdot\delta(O_1) \geqslant (1-\beta_1)\cdot\alpha\cdot\opt$ edges. Observe also that, by~(\ref{mu}) and~(\ref{1+mu}), $\nicefrac{k}{k_1} = \nicefrac{1+\mu}{\mu}$. In any of the remaining $\nicefrac{1+\mu}{\mu} - 2 = \nicefrac{1-\mu}{\mu}$ clusters, their vertices obviously cover more than $\zeta\cdot\delta(O_1) = \zeta\cdot\alpha\cdot\opt$ edges (indeed, by the average argument of Claim~\ref{fact1}, more than $k_1\cdot\nicefrac{\zeta\cdot\delta(O_1)}{k'_1} \geqslant \zeta\cdot\delta(O_1))$). We so have:
$$
\left|\SOL_4\right| \geqslant \left[\left(2 - \beta_1\right) + \frac{1 - \mu}{\mu}\cdot\zeta\right]\cdot\delta\left(O_1\right) = \left[\left(2 - \beta_1\right) + \frac{1 - \mu}{\mu}\cdot\zeta\right]\cdot\alpha\cdot\opt
$$
that proves~(\ref{deka}).

Let us now get some more insight in the value of~$\SOL_4$. By extending the discussion just above, the~$k_1$ vertices of cluster~$i$ will cover more than:
\begin{equation}\label{pii}
\frac{k_1}{k_1 - \sum\limits_{j=1}^{i-1}\kappa_j}\cdot\left(\sum_{j=i}^{\pi}\beta_j + \zeta\right)\cdot\delta\left(O_1\right) \geqslant \left(1 - \sum_{j=1}^{i-1}\beta_j\right)\cdot\delta\left(O_1\right)
\end{equation}
Furthermore, as seen previously, all clusters below the~$\pi$ first ones containing the $k_1-k'_1$ captured vertices of~$O_1$, will cover more than~$\zeta\cdot\delta(O_1)$ each.

Hence, summing~(\ref{pii}) for $i=1$ to~$\pi$, taking into account the remark just above, and setting $\beta_0 = 0$, the following holds:
\begin{eqnarray}\label{star1}
\left|\SOL_4\right| &\geqslant& \left(\sum_{i=0}^{\pi-1}\left(1 - \sum_{j=0}^{i}\beta_j\right) + \left(\frac{1+\mu}{\mu}-\pi\right)\cdot\zeta\right)\cdot\delta\left(O_1\right) \nonumber \\
&=& \left(\pi - \sum_{i=1}^{\pi-1}(\pi-i)\cdot\beta_i + \left(\frac{1+\mu}{\mu}-\pi\right)\cdot\zeta\right)\cdot\delta\left(O_1\right) \nonumber \\
&=& \left(\pi - \pi\cdot\sum_{i=1}^{\pi-1}\beta_i + \sum_{i=1}^{\pi-1}i\cdot\beta_i + \left(\frac{1+\mu}{\mu}-\pi\right)\cdot\zeta\right)\cdot\delta\left(O_1\right)
\end{eqnarray}
Observe now that:
\begin{equation}\label{star2}
\pi\cdot\sum_{i=1}^{\pi-1}\beta_i = \pi\cdot\left(1 - \beta_{\pi} - \zeta\right)\cdot\delta\left(O_1\right)
\end{equation}
and combine~(\ref{star2}) with~(\ref{star1}). Then, the latter becomes:
\begin{eqnarray}\label{star3}
\left|\SOL_4\right| &\geqslant& \left(\pi - \pi\cdot\left(1 - \beta_{\pi} - \zeta\right) + \sum_{i=1}^{\pi-1}i\cdot\beta_i + \left(\frac{1+\mu}{\mu}-\pi\right)\cdot\zeta\right)\cdot\delta\left(O_1\right) \nonumber \\
&=& \left(\sum_{i=1}^{\pi}i\cdot\beta_i + \pi\zeta +  \left(\frac{1+\mu}{\mu}-\pi\right)\cdot\zeta\right)\cdot\delta\left(O_1\right) \nonumber \\
&=& \left(\sum_{i=1}^{\pi}i\cdot\beta_i + \left(\frac{1+\mu}{\mu}\right)\cdot\zeta\right)\cdot\delta\left(O_1\right) \nonumber \\
&=& \left(\sum_{i=1}^{\pi}\beta_i + \sum_{i=2}^{\pi}\beta_i + \sum_{i=3}^{\pi}(i-2)\cdot\beta_i + \frac{1+\mu}{\mu}\cdot\zeta\right)\cdot\delta\left(O_1\right) \nonumber \\
&=& \left(\left(1 - \zeta\right) + \left(1 - \beta_{1} - \zeta\right) + \sum_{i=3}^{\pi}(i-2)\cdot\beta_i + \frac{1+\mu}{\mu}\cdot\zeta\right)\cdot\delta\left(O_1\right) \nonumber \\
&=& \left(\left(2-\beta_1\right) + \sum_{i=3}^{\pi}(i-2)\cdot\beta_i + \frac{1-\mu}{\mu}\cdot\zeta\right)\cdot\delta\left(O_1\right) \nonumber \\
&\geqslant& \left(\left(2-\beta_1\right) + \sum_{i=3}^{\pi}\cdot\beta_i + \frac{1-\mu}{\mu}\cdot\zeta\right)\cdot\delta\left(O_1\right) 
\end{eqnarray}
Set $\sum_{i=3}^{\pi}\cdot\beta_i\cdot\delta(O_1) = X$. These edges are covered by both~$O_1$ and~$\SOL_4$. Then,~(\ref{star3}) becomes:
\begin{equation}\label{star3'}
\left|\SOL_4\right| \geqslant \left(\left(2-\beta_1\right) + \frac{1-\mu}{\mu}\cdot\zeta\right)\cdot\delta\left(O_1\right) + X = \left(\left(2-\beta_1\right) + \frac{1-\mu}{\mu}\cdot\zeta\right)\cdot\alpha\cdot\opt + X
\end{equation}
On the other hand, consider Item~\ref{2k1} in~$\SOL_4$. The~$2\cdot k_1$ best vertices from~$V_1$ cover $(1 - \zeta)\cdot\delta(O_1) - X$ edges of the optimum. Let $Y + [(1 - \zeta)\cdot\delta(O_1) - X]$ be the total number of edges covered by those vertices. Then, best $k-2\cdot k_1$ vertices of~$V_2$ will cover at least as many edges as the $k-2\cdot k_1$ best vertices of~$O_2$, that will cover at least $(\nicefrac{(k-2\cdot k_1)}{k_2})\cdot\delta'(O_2) - Y$. Putting all this together, we get:
\begin{eqnarray}\label{star4}
\left|\SOL_4\right| &\geqslant& Y + \left[(1 - \zeta)\delta(O_1) - X\right] + \frac{k-2\cdot k_1}{k_2}\cdot\delta'\left(O_2\right) - Y \nonumber \\
&=& \left((1 - \zeta)\cdot\alpha + \frac{k-2\cdot k_1}{k_2}\cdot(1-\alpha)\right)\cdot\opt - X \nonumber \\
&{(\ref{mu}),(\ref{1+mu}) \atop =}& \left((1 - \zeta)\cdot\alpha + (1-\mu)\cdot(1-\alpha)\right)\cdot\opt - X
\end{eqnarray}
Expression~(\ref{star3'}) is increasing with~$X$, while~(\ref{star4}) is decreasing. Equality of them, leads after some easy algebra to:
\begin{equation}\label{x}
X = \left(\frac{(1-\mu)-\alpha\cdot(2-\mu)+\beta_1\cdot\alpha}{2} - \frac{\zeta\cdot\alpha}{2\cdot\mu}\right)\cdot\opt
\end{equation}
Embedding~(\ref{x}) to~(\ref{star3'}) and dividing the ratio obtained by~$\opt$, derives the ratio claimed by~(\ref{star}).~\eprf

\section{\ldots and its approximation ratio}\label{approxratiosec}

The objective of this section is to prove the following theorem.
\thm\label{thethm}
\mkvc{} is combinatorially approximable within ratio~0.7.
\ethm
\prf
For the proof we propose an exhaustive parameter-elimination method (very probably non-optimal) that has the advantage to be quite simple. It consists of subsequently eliminating parameters from the ratios proved in Lemmata~\ref{sol1lem} to~\ref{sol4lem} until two ratios that are only functions of~$\mu$ are got. These ratios have opposite monotonies with respect to this parameter, hence by equalizing them we determine a lower bound for the overall ratio of the algorithm.

\subsubsection*{Elimination of~$\theta$: ratios~(\ref{pente}) and~(\ref{exi})}

Equalizing ratios given by~(\ref{pente}) and~(\ref{exi}) leads to $2\alpha\cdot\theta = \alpha - \lambda\cdot(1-\alpha) \Rightarrow \alpha\cdot\theta = \nicefrac{\alpha - \lambda\cdot(1-\alpha}{2}$ and embedding it in~(\ref{pente}) derives:
\begin{equation}\label{dwdeka}
\frac{2-\alpha\cdot(1-\lambda)-\lambda}{2}
\end{equation}

\subsubsection*{Elimination of~$\lambda$: ratios~(\ref{epta}) and~(\ref{dwdeka})}

Equalizing ratios given by~(\ref{epta}) and~(\ref{dwdeka}) gives $\lambda = \nicefrac{\alpha}{(1-\alpha)\cdot(1+2\cdot\mu)}$. This, together with~(\ref{epta}), derives:
\begin{equation}\label{dekatria}
1 - \alpha\cdot\frac{1+\mu}{1+2\cdot\mu}
\end{equation}

\subsubsection*{Elimination of~$\gamma$: ratios~(\ref{tessera}) and~(\ref{ennia})}

It gives $\gamma = \nicefrac{1-\alpha-\zeta\cdot\alpha}{2\cdot(1-\alpha)}$ and the ratio obtained is:
\begin{equation}\label{dekatessera}
\frac{1+\alpha-\zeta\cdot\alpha}{2}
\end{equation}

\subsubsection*{Elimination of~$\zeta$: ratios~(\ref{deka}) and~(\ref{dekatessera})}

We have:
\begin{eqnarray}\label{zeta}
& & \left(2 - \beta_1\right)\cdot\alpha + \frac{(1-\mu)\cdot\zeta\cdot\alpha}{\mu} = \frac{1+\alpha-\zeta\cdot\alpha}{2} \nonumber \\
&\Rightarrow& \frac{2-\mu}{\mu}\cdot\zeta\cdot\alpha = 1 + \alpha -2\cdot\left(2-\beta_1\right)\cdot\alpha = 1 - 3\cdot\alpha+2\cdot\beta_1\cdot\alpha \nonumber \\
&\Rightarrow& \zeta = \frac{\mu}{2-\mu}\cdot\frac{1-3\cdot\alpha+2\cdot\beta_1\cdot\alpha}{\alpha}
\end{eqnarray}
and embedding~(\ref{zeta}) in~(\ref{deka}), we get:
\begin{eqnarray}\label{dekapente}
& & \left(2 - \beta_1\right)\cdot\alpha + \frac{1-\mu}{2-\mu}\cdot\left(1-\alpha\cdot\left(3-2\cdot\beta_1\right)\right) \nonumber \\
&=& \frac{(2-\mu)\cdot\left(2 - \beta_1\right)\cdot\alpha + (1-\mu) - \alpha\cdot(1-\mu)\cdot\left(3-2\cdot\beta_1\right)}{2-\mu} \nonumber \\
&=& \frac{(1-\mu) + \alpha\cdot\left(1+\mu-\mu\cdot\beta_1\right)}{2-\mu}
\end{eqnarray}

\subsubsection*{First elimination of~$\beta_1$: ratios~(\ref{ena}) and~(\ref{dekapente})}

We have:
\begin{eqnarray}\label{beta1}
& & 1 - \alpha + \beta_1\cdot\alpha = \frac{(1-\mu) + \alpha\cdot\left(1+\mu-\mu\cdot\beta_1\right)}{2-\mu} \nonumber \\
&\Rightarrow& 2\cdot\beta_1\cdot\alpha = (1-\mu) + \alpha\cdot(1+\mu) - (2-\mu) + \alpha\cdot(2-\mu) = -1 + 3\cdot\alpha \nonumber \\
&\Rightarrow& \beta_1 = \frac{3\cdot\alpha - 1}{2\cdot\alpha}
\end{eqnarray}
Now, combination of~(\ref{ena}) and~(\ref{beta1}) derives:
\begin{equation}\label{1+alpha2}
1-\alpha+\beta_1\cdot\alpha = 1-\alpha+\frac{3\cdot\alpha - 1}{2} = \frac{1+\alpha}{2}
\end{equation}

\subsubsection*{First ratio function of~$\mu$: combination of ratios~(\ref{dekatria}) and~(\ref{1+alpha2})}

Ratio~(\ref{dekatria}) is decreasing with~$\alpha$, while ratio~(\ref{1+alpha2}) is increasing. Combination of them allows elimination~$\alpha$ in order to get a first ratio that is only a function of~$\mu$. Equalizing~(\ref{dekatria}) and~(\ref{1+alpha2}) gives:
\begin{eqnarray}\label{alpha}
& & 1 - \alpha\cdot\frac{1+\mu}{1+2\cdot\mu} = \frac{1+\alpha}{2} \Rightarrow \alpha\cdot(4\cdot\mu+3) = 2\cdot\mu +1 \nonumber \\
&\Rightarrow& \alpha = \frac{2\cdot\mu+1}{4\cdot\mu+3}
\end{eqnarray}
and embedding~(\ref{alpha}) in~(\ref{dekatria}) derives:
\begin{eqnarray}\label{Ratio1}
1 - \alpha\cdot\frac{1+\mu}{1+2\cdot\mu} &=& 1 - \frac{2\cdot\mu+1}{4\cdot\mu+3}\cdot\frac{1+\mu}{1+2\cdot\mu} \nonumber \\
&=& \frac{2+3\cdot\mu}{3+4\cdot\mu}
\end{eqnarray}

\subsubsection*{Second elimination of~$\beta_1$: ratios~(\ref{ena}) and~(\ref{star})}

Revisit ratio~(\ref{star}) and observe that its last term~$\nicefrac{(1-2\cdot\mu)\cdot\zeta\cdot\alpha}{2\cdot\mu}$ is negative when $\mu \geqslant \nicefrac{1}{2}$. On the other hand, ratio~(\ref{Ratio1}) is increasing with~$\mu$ and bounded below by~0.7 as long as $\mu \geqslant \nicefrac{1}{2}$. We so seek an ``interesting'' ratio when $\mu \leqslant \nicefrac{1}{2}$ and, in this case $\nicefrac{(1-2\cdot\mu)\cdot\zeta\cdot\alpha}{2\cdot\mu} \geqslant 0$ and can be omitted.

Hence, combination of ratios~(\ref{ena}) and~(\ref{star}), for $\mu \leqslant \nicefrac{1}{2}$, leads to:
\begin{eqnarray}\label{beta1bis} 
& & \frac{(1-\mu)+\alpha\cdot(2+\mu)-\alpha\cdot\beta_1}{2} = 1-\alpha+\beta_1\cdot\alpha \nonumber \\
&\Rightarrow& (1-\mu)+\alpha\cdot(2+\mu)-\alpha\cdot\beta_1 = 2 - 2\cdot\alpha + 2\cdot\beta_1\cdot\alpha \nonumber \\
&\Rightarrow& 3\cdot\beta_1\cdot\alpha = -(1+\mu) + \alpha\cdot(4+\mu) \Rightarrow \beta_1 = \frac{\alpha\cdot(4+\mu)-(1+\mu)}{3\cdot\alpha}
\end{eqnarray}
Then, combining~(\ref{ena}) and~(\ref{beta1bis}), derives this time:
\begin{equation}\label{2trita-alpha}
1-\alpha+\beta_1\cdot\alpha = \frac{3-3\cdot\alpha + \alpha\cdot(4+\mu)-(1+\mu)}{3} = \frac{2-\mu + \alpha\cdot(1+\mu)}{3}
\end{equation}

\subsubsection*{Second ratio function of~$\mu$: combination of ratios~(\ref{2trita-alpha}) and~(\ref{dekatria})}

Once again, ratio~(\ref{dekatria}) is decreasing with~$\alpha$, while ratio~(\ref{2trita-alpha}) is increasing. Combination of them allows elimination~$\alpha$ in order to get a second ratio exclusively function of~$\mu$. Equalizing~(\ref{dekatria}) and~(\ref{2trita-alpha}) gives:
\begin{eqnarray}\label{alphabis}
& & 1 - \alpha\cdot\frac{1+\mu}{1+2\cdot\mu} = \frac{2-\mu + \alpha\cdot(1+\mu)}{3} \Rightarrow 2\cdot\alpha\cdot\frac{(1+\mu)\cdot(2+\mu)}{3\cdot(1+2\cdot\mu)} = \frac{1+\mu}{3} \nonumber \\
&\Rightarrow& \alpha = \frac{1+2\cdot\mu}{2\cdot(2+\mu)}
\end{eqnarray}
and embedding~(\ref{alphabis}) in~(\ref{dekatria}) derives:
\begin{equation}\label{Ratio2}
1 - \alpha\cdot\frac{1+\mu}{1+2\cdot\mu} = \frac{3+\mu}{4+2\cdot\mu}
\end{equation}

\subsubsection*{Final ratio}

As noted above, ratio~(\ref{Ratio1}) increases with~$\mu$, while~(\ref{Ratio2}) decreases. The value of~$\mu$ guaranteeing equality of these ratios also gives a lower bound for them. This value is $\mu = \nicefrac{1}{2}$ and, with this value, both ratios become 0.7. \eprf

\bigskip

\noindent
\textbf{Acknowledgement.} The very fruitful and stimulating discussions I had with Edouard Bonnet and Georgios Stamoulis are gratefully acknowledged. Many thanks to Edouard Bonnet for his very pertinent comments on a first draft of the paper.


\end{document}